\documentclass[12pt]{iopart}
\pdfoutput=1

\usepackage{graphicx}
\usepackage{float}
\begin{document}

\title[Temperature dependence of the $(\pi,0)$ anomaly in the 2DQHAFSL]{Temperature dependence of the $(\pi,0)$ anomaly in the excitation spectrum of the 2D quantum Heisenberg antiferromagnet}

\author{W. Wan$^1$, N.B. Christensen$^1$, A.W. Sandvik$^{2,3}$, P. Tregenna-Piggott$^4$, G.J. Nilsen$^5$, M. Mourigal$^6$, T.G. Perring$^5$, C.D. Frost$^5$, D.F. McMorrow$^7$, H.M. R\o nnow$^8$}
\address{$^1$Department of Physics, Technical University of Denmark, DK-2800 Kongens Lyngby, Denmark}
\address{$^2$Department of Physics, Boston University, 590 Commonwealth Avenue, Boston, Massachusetts 02215, USA}
\address{$^3$Beijing National Laboratory for Condensed Matter Physics and Institute of Physics,
Chinese Academy of Sciences, Beijing 100190, China}
\address{$^4$Laboratory for Neutron Scattering and Imaging, Paul Scherrer Institut, CH-5232 Villigen PSI, Switzerland}
\address{$^5$ISIS Facility, STFC Rutherford-Appleton Laboratory, Didcot OX11 0OX, UK}
\address{$^6$School of Physics, Georgia Institute of Technology, 837 State Street, Atlanta, Georgia 30332, USA}
\address{$7$London Centre for Nanotechnology and Department of Physics and Astronomy,
University College London, London WC1E 6BT, United Kingdom}
\address{$^8$Laboratory for Quantum Magnetism, Institute of Physics, Ecole Polytechnique F\'ederale de Lausanne (EPFL), CH-1015 Lausanne, Switzerland}

\ead{nbch@fysik.dtu.dk}
\date{\today}

\begin{abstract}
It is well established that in the low-temperature limit, the two-dimensional quantum Heisenberg antiferromagnet on a square lattice (2DQHAFSL) exhibits an anomaly 
in its spectrum at short-wavelengths on the zone-boundary. In the vicinity of the $(\pi,0)$ point the pole in the one-magnon response exhibits a downward dispersion, 
is heavily damped and attenuated, giving way to an isotropic continuum of excitations extending to high energies. The origin of the anomaly and 
the presence of the continuum are of current theoretical interest, with suggestions focused around the idea that the latter evidences the existence of 
spinons in a two-dimensional system. Here we present the results of neutron inelastic scattering experiments and Quantum Monte Carlo calculations on the metallo-organic compound 
Cu(DCOO)$_2\cdot 4$D$_2$O (CFTD), an excellent physical realisation of the 2DQHAFSL, designed to investigate how the anomaly at $(\pi,0)$ evolves up to 
finite temperatures $T/J\sim2/3$. Our data reveal that on warming the anomaly survives the loss of long-range, three-dimensional order, and is thus a robust 
feature of the two-dimensional system. With further increase of temperature the zone-boundary response gradually softens and broadens, washing out the $(\pi,0)$ anomaly. 
This is confirmed by a comparison of our data with the results of finite-temperature Quantum Monte Carlo simulations where the two are found to be in good accord. 
At lower energies, in the vicinity of the antiferromagnetic zone centre, there was no significant softening of the magnetic 
excitations over the range of temperatures investigated.
\end{abstract}
\maketitle

\section{Introduction}

\subsection{Roger Cowley's interests in phase transitions in magnetic materials}

Throughout his career, Roger Cowley maintained a keen interest in the magnetic properties of materials, most especially their magnetic structures and excitations that manifest the underlying interactions. Roger's experimental technique of choice was neutron scattering, a field which he played a leading role in developing, often backed up by original theoretical contributions. Threading through this work was a focus on the manifold aspects of the magnetism of insulators. These are of enduring interest because of their great flexibility in terms of being able to tailor a material to correspond closely to a particular model Hamiltonian through chemical means. Thus by judicious choice of magnetic ion it is possible to realise materials with Ising, XY or Heisenberg (and more recently Kitaev-Heisenberg) spin symmetry, and to place the ions on lattices with one, two or three-dimensional connectivities. Roger used such materials as platforms to both test and develop key theories in the physics of phase transitions, 

Roger's work on magnetic insulators started in the late 1960s, with a notable early milestone being the observation in CoF$_2$ of the first magnetic exciton \cite{cowley1967}. His research in this early period expanded to include the study of the effects of substitutional disorder on antiferromagnets. Indeed, already by 1972 he and Bill Buyers were able to produce an authoritative review of the theoretical and experimental achievements in that field up to that time \cite{cowley1972}. The study of substitutionally disordered antiferromagnets  was an abiding theme of Roger's research over the next decade or so. Working with Bob Birgeneau, Gen Shirane and others, Roger produced a series of landmark papers that addressed the nature of the percolation transition in dilute systems by studying a series of magnetic compounds with both magnetic and non-magnetic impurities
\cite{cowley1977,birgeneau1980,cowley1980b}. A little later, the same team went on to investigate the nature of the phase transition in the random field Ising model \cite{yoshizawa1982,birgeneau1983}. This model assumed a particular prominence at the time for understanding the effect of disorder on phase transitions with wide-spread ramifications, such as dimensional reduction: the prediction that systems with random fields behave like the pure system in two fewer dimensions \cite{imry1975}. Moreover, the realisation that the random field Ising model maps exactly onto a site random Ising magnet in a uniform magnetic field, opened the door to experiments in which it was possible to continuously tune the effects of disorder by changing the applied field. Although many of the predictions of theory were substantiated by Roger's experiments, others were not, with open issues remaining to this day \cite{Fytas2018}.
This period through to the mid 1980s may be characterised as a golden age for the fields of phase transitions and statistical mechanics, culminating in the award of the 1982 Nobel Prize in Physics to Kenneth Wilson for his development of the renormalisation group. It is against this backdrop that Roger's outstanding contributions to these fields should be considered.   

All of the experiments described above were performed on triple-axis spectrometers at reactor sources. The availability from the mid 1980s onwards of spallation sources dedicated to the production of neutrons, such as ISIS in the UK, presented new challenges and opportunities for the study of magnetic excitations. The challenges were not insignificant at the time, requiring a paradigm shift in how to perform neutron scattering experiments optimally using a pulsed source. Instead of collecting data in constant wavevector $\mathbf{Q}$ or energy transfer ${\cal E}$ scans typically acquired with a 
triple-axis spectrometer, experiments using a pulsed source  produce (large) time-of-flight, volumetric $(\mathbf{Q},{\cal E}=\hbar\omega)$ data sets, from which scans along arbitrary trajectories can be constructed in software. The opportunities were the other side of the coin, as the  coverage of $(\mathbf{Q},{\cal E})$ enabled the visualisation of the complete excitation spectrum at all relevant $\mathbf{Q}$ and ${\cal E}$. Roger was firmly in the vanguard of these developments. Here his work on low-dimensional Heisenberg antiferromagnets (HAF) is of most relevance to the current article. Along with Alan Tennant and Stephen Nagler, he  showed definitively that the spin waves in the 1D quantum ($S$=1/2) HAF fractionalise into spinons in agreement with theory \cite{tennant1993}. 
It is our opinion that this study, as much as anything else, served to establish the credibility of time-of-flight techniques for the study of magnetic excitations in single crystals, helping to convince what had up until that point been a somewhat sceptical community. One of the last projects Roger worked on at ISIS was to investigate the nature of the magnetic excitations in  the 2D classical (large $S$), square lattice antiferromagnet Rb$_2$MnF$_4$ \cite{huberman2005}, and their evolution with temperature up to $k_BT\simeq$4$JS$  \cite{huberman2008} where $J$ is the nearest neighbour 2D exchange constant. The complete $(\mathbf{Q},{\cal E})$  coverage afforded by the technique provided 
data across the entire Brillouin zone which were found to be in excellent agreement with the results of classical simulations.

\subsection{The 2D quantum Heisenberg antiferromagnet on the square lattice}

The remainder of this article is concerned with the 2D quantum ($S$=1/2) HAF on the square-lattice (2DQHAFSL) with only nearest neighbor interactions.
Unlike its classical counterpart this model does exhibit explicit quantum effects. Whereas no formal proof exists, numerical evidence unanimously indicates that the 2DQHAFSL orders only at $T=0$ \cite{RevModPhys.63.1}. The ordering wavevector is $(\pi, \pi)$ and the sublattice magnetization is reduced to 60$\%$ of its classical expectation value $\mu \simeq 0.6\mu_{B}$. Physical realizations of the 2DHAFSL model include weak interlayer couplings $J'$, the effect of which is amplified upon cooling by the divergent in-plane correlations length $\xi(T)$. The approximate mean-field ordering condition $k_BT_N=zJ'S(S+1)[\xi(T_N)/a]^2$, where $z$ is the in-plane coordination number, and $a$ is the nearest neighbor distance, has been refined and extended to higher spatial dimensions \cite{PhysRevB.74.184407}.

The low-energy spin fluctuations of the ordered state are empirically known to be well-described by classical spin wave theory, modified by a multiplicative quantum renormalization factor $Z_c=1.18$ for the spin wave velocity. The short wavelength spin excitations, however, are strongly influenced by quantum fluctuations. In particular, detailed investigations of the spin dynamics along the magnetic Brillouin zone boundary, connecting $(\pi, 0)$ to $(\pi/2, \pi/2)$ exhibit several anomalies: 
(i) $7\%$ downwards dispersion at $(\pi, 0)$, (ii) about $50\%$ loss of spectral weight in the spin-wave peak at $(\pi, 0)$, and (iii) significant continuum above the spin-wave peak occurring in both transverse and longitudinal fluctuation channels such that the continuum appears spin-isotropic at $(\pi, 0)$ even in the symmetry-broken antiferromagnetically ordered state.
One or more of these anomalies at $(\pi, 0)$  have been observed in multiple physical realizations of the 2DQHAFSL model including the oxychlorides Sr$_2$Cu$_3$O$_4$Cl$_2$, Ba$_2$Cu$_3$O$_4$Cl$_2$ and Sr$_2$CuO$_2$Cl$_2$ \cite{PhysRevLett.83.852,PhysRevB.96.014410,PhysRevB.89.180410}, Copper deuteroformate tetradeuterate (Cu(DCOO)$_2\cdot 4$D$_2$O, henceforth abbreviated CFTD) \cite{PhysRevLett.87.037202,Christensen15264,dalla2015fractional} and Cu(pz)$_2$(ClO$_4$)$_2$ \cite{PhysRevB.81.134409}. The missing weight in the spin-wave pole and increased continuum above at $(\pi,0)$ was also observed in the undoped cuprate La$_2$CuO$_4$\cite{PhysRevLett.105.247001}, whereas the zone-boundary dispersion in the cuprates is inversed due to 4-spin interactions steming from the Hubbard model heritage of the Heisenberg Hamiltonian \cite{Guarise2010,DallaPiazza2012}. Among these materials, the most complete studies have been performed on CFTD, where polarized neutrons were exploited to distinguish magnetic from phonon scattering, and to separate the transverse and longitudinal components of the spin excitation spectrum in the magnetically ordered state.

It has been known for some time that spin wave theory -- even when taken to second \cite{PhysRevB.72.014403} or third \cite{Syromyatnikov_2010} order in the expansion parameter $1/S$  -- is not capable of fully reproducing the zone-boundary dispersion of the spin excitations in square lattice antiferromagnets. 
Series expansion from the Ising limit reproduced the dispersion, but has not addressed the magnon spectral weight or the continuum. Exact diagonalization can reproduce the trends \cite{PhysRevB.79.195102,PhysRevB.99.054432}, but due to finite size limitations cannot be used for quantitative comparison to experiments. Quantum Monte Carlo (QMC) reproduces the zone boundary dispersion and within the uncertainties tied to using analytic continuation can reproduce the spectal weight and lineshape anomalies \cite{Sylju_sen_2000,PhysRevLett.86.528}. However these numerical methods do not in themselves shed light on the underlying nature of the anomalies.

An approach based on Gutzwiller projected particle-hole excitations qualitatively reproduced the $(\pi, 0)$ anomalies with states that upon inspection corresponded to delocalized spin-flips - 2D equivalents of the spinon pairs \cite{dalla2015fractional,PhysRevB.98.100405} demonstrated by Roger Cowley and his collaborators in the 1D case \cite{tennant1993}. In this picture, the spin-wave excitation correspond to bound spinon pairs, which deconfine around $(\pi, 0)$ but not at $(\pi/2,\pi/2)$. 
The notion of nearly deconfined spinons was also supported by a recent QMC study \cite{PhysRevX.7.041072}, which included a demonstration that an effective model of interacting magnons and spinons is able to reproduce both the zone boundary dispersion and the decreased spectral weight of the magnon mode at $(\pi, 0)$. The underlying picture of a 2D spinon continuum with a lower boundary that can fall below the pure magnon energy at $(\pi, 0)$ but does not do so at $(\pi/2, \pi/2)$ has some resemblance to an RPA calculation of the transverse part of the excitation spectrum in an RVB treatment of the square lattice \cite{PhysRevLett.86.1626}.

In contrast to the Gutzwiller, QMC and coupled magnon-spinon approaches discussed above, the $(\pi, 0)$ anomalies have also been reproduced by continuous unitary transformation methods \cite{PhysRevLett.115.207202}, which rather represent interacting multi-spin-wave states. Given that this approach could yield even quantitative agreement with experimental lineshapes, it could perhaps seem a preferred picture, but could also be construed \cite{PhysRevX.7.041072} as merely an effective model neglecting a more complicated reality of magnons as confined spinons. More generally, the 2DQHAFSL model and its physical realization simply have momentum and energy dependent spectral response functions, which do not even need to be describable by any quasi-particle picture. It may be that some features are best described as multi-spin-waves, others as confined and deconfined spinons, and still others are not described by either.

In this context, it is interesting to investigate how the spectrum evolves as thermal fluctuations destroy long range antiferromagnetic N\'eel order and lead to gradually shorter correlation lengths \cite{PhysRevLett.82.3152}. In a spinon-picture, the lack of N\'eel order would effectively allow spinons to deconfine. In a spin-wave picture, single spin-waves would acquire finite life-time resulting in symmetric broadening of the line-shape, but accompanied by a possible change also in the multi-spin-wave continuum, which could account for any asymmetry in the line shapes. While waiting for theoretical predictions based on either picture, we here report an experimental investigation of the finite temperature excitation spectrum of the 2D $S=1/2$ square lattice Heisenberg antiferromagnet. The results we present touch not only on the zone boundary anomalies, but also on the longer wavelength excitations, studied by Roger Cowley in the classical spin case \cite{huberman2008}, and more recently by Resonant Inelastic X-ray Scattering (RIXS), where it was found that upon doping, the dispersive excitations soften much more drastically along the $(\pi, \pi)$ direction than along the $(\pi, 0)$ direction \cite{guarise2014}. 

\section{Experimental and computational techniques}

\subsection{Neutron scattering}
To address the issues raised above, we performed a neutron scattering experiment aimed at studying the temperature dependence of the magnetic excitations in CFTD. This transparent blue insulating material is among the best realizations of the 2D square lattice quantum $S=1/2$ Heisenberg model. 

At room temperature, CFTD is monoclinic (space group P2$_1$/a), but under cooling undergoes an antiferroelectric transition at $246$\ K, which involves a doubling of the unit cell along the crystallographic $c$ axis. At $100$\ K the space group is P2$_1$/n (non-standard setting of space group 14) with lattice constants $a=8.113$ \AA, $b=8.119$ \AA, $c=12.45$ \AA, and $\beta=100^\circ$. Within the $ab$-plane, the Cu$^{2+}$ ions are coordinated by four oxygens from the formate molecules and form an almost ideal square lattice arrangement with nearest neighbor lattice constant $5.739$ \AA. The octahedral coordination is completed by oxygens from the crystal bound water molecules between the planes. 

The Curie-Weiss temperature of CFTD is $\Theta=-175$ K \cite{JR9590001359}, and three dimensional antiferromagnetic order sets in below $T_N=$16.5 K. The staggered magnetic moment $0.48(2)$ $\mu_B$ in the ordered state was determined by Burger {\it et al} \cite{Burger1980} who found the spins to be rotated $8^{\circ}$ degrees away from the $a$ axis towards the $c$ axis. Prior inelastic neutron scattering experiments have found an antiferromagnetic exchange coupling $J=6.19(2)$ meV \cite{Christensen15264} between nearest neighbors, and no evidence for longer range interations within the copper-formate planes. By analysing magnetization data, Yamagata and coworkers \cite{Yamagata1980,Yamagata1981} determined the ratio $J_c/J$ of the interlayer exchange $J_c$ to the intraplane interaction, $J$, to be of order $10^{-5}$. Their analysis also allowed for a Dzyaloshinskii-Moriya (DM) interaction with the DM vector close to $c$ and of magnitude 0.05J. 

The sample used in the inelastic neutron scattering experiment reported here consisted of three co-aligned high-quality solution grown single crystals (For details of the growth procedure, see Ref. \cite{dalla2015fractional}, supplementary information). The total weight of the crystals was approximately 12 grams. Exploiting the two-dimensional nature of the spin correlations in CFTD, the crystallographic $c$-axis was aligned along the neutron beam of the MAPS spectrometer at the ISIS Facility, STFC Rutherford-Appleton Laboratory. With an incident neutron energy of $36.25$ meV and a Fermi chopper frequency of $200$ Hz, we obtained an energy resolution of 1.3 meV (FWHM) at the energy of the zone boundary magnons, $2Z_cJ \simeq 14.5$ meV. With this instrument configuration, we were able to probe the full dynamic spin correlation function $S({\bf Q},\omega)$, where ${\bf Q}$ is the projection of the neutron momentum transfer perpendicular to the crystallographic $c$-axis. Data was collected at 6, 20, 35, 40, 50, and 80 K, corresponding to $T/J=0.08$, $0.28$, $0.49$, $0.56$, $0.70$, and $1.11$. 

\subsection{Quantum Monte Carlo}

The numerical results for $S(\mathbf Q,\omega)$ were obtained by stochastic analytic continuation
\cite{sandvik1998,syljuasen2008,PhysRevX.7.041072} of imaginary-time correlation functions $G(\mathbf{Q},\tau)$ computed with the stochastic series expansion QMC method, implemented according to Ref. \cite{sandvik2010}. The correlation functions were computed on an imaginary-time grid with spacing $\Delta_\tau/J = 0.01$, using the exact time-slicing method discussed in Ref. \cite{sandvik2019}. 
The lattices sizes ranged from 32$\times$32 for $T/J=2/3$ to 128$\times$128 for $T/J=1/4$, 
and tests for other sizes indicate that no significant finite-size effects are left for the $\mathbf Q$ 
points considered here. The individual error bars on the imaginary-time data, after normalizing such that 
$G(\mathbf Q,\tau = 0)$ = 1, were less than 10$^{-5}$ in all cases. 
All covariance effects were properly accounted for in the analytic continuation. 
We used the recently introduced parametrization of the spectrum by a number $N_\omega$ 
of equal-amplitude $\delta$-functions (here with $N_\omega$ = 5000), whose frequencies $\omega_i$, $i$ = 1,$...$,$N_\omega$, 
were sampled under conditions guaranteeing a good fit to the imaginary-time data while at the same time avoiding overfitting. These computational procedures are discussed in detail in Ref. \cite{PhysRevX.7.041072}. 
The mean density of $\delta$-functions in the sampling process for given $\mathbf{Q}$ 
is accumulated in a histogram and represents the final result for $S(\mathbf Q,\omega)$.

Numerical analytic continuation is always associated with uncertainties, and there are known limitations on the frequency resolution. In the present case, we do not expect any sharp spectral features at the high temperatures for which we carry out comparisons, in contrast to the $T$=0 case considered in Ref. \cite{PhysRevX.7.041072}, where special procedures were developed to resolve the $\delta$-function contribution arising from the single-magnon pole. Tests with synthetic data with the same noise level as in the QMC data indicate that broadening and other spectral distortions should be rather mild in the results presented here. We have not carried out any further Gaussian convolution to account for the instrumental resolution.

\section{Results and analysis}

\subsection{Temperature dependence of the excitation spectrum across the Brillouin zone}

\begin{figure}[H]
\centering
    \includegraphics[width=0.8\textwidth]{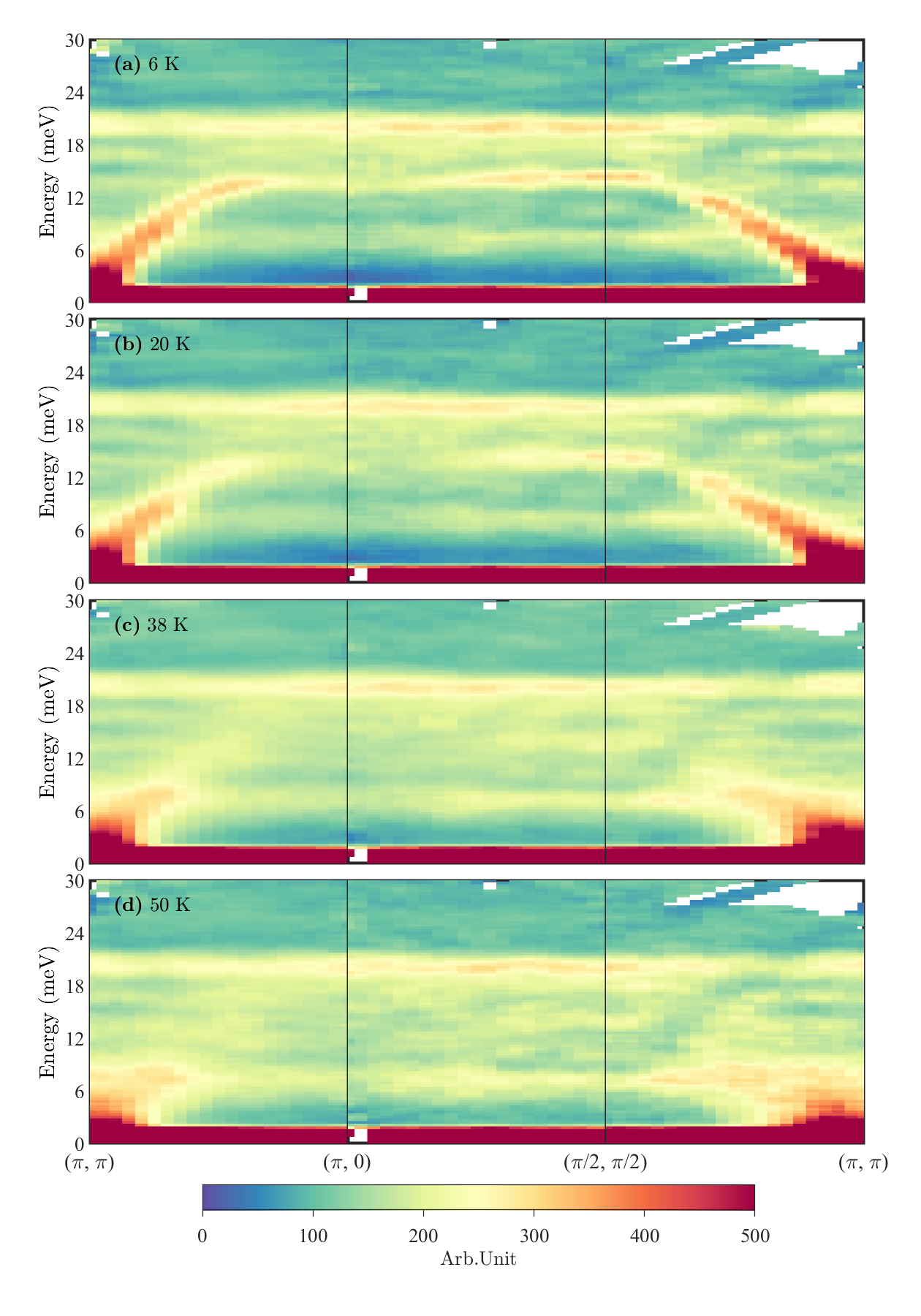}
    \caption{Overview of the magnetic and lattice excitation spectra of CFTD at (a) 6~K, (b) 20~K, (c) 38~K and (d) 50~K (The 38~K data set in panel (c) is actually the average of raw data obtained at 35~K and 40~K) measured by time-of-flight inelastic neutron scattering. The momentum axis follows the standard path around the Brillouin zone of the square lattice. Intensities are shown in arbitrary units.}
    \label{fig_NS_raw}
\end{figure}

Figure\ \ref{fig_NS_raw} illustrates the evolution of the spin and lattice excitations along high-symmetry directions in the Brillouin zone at temperatures below and above the N\'eel temperature, $T_N=$16.5 K. To produce these plots, the raw data were averaged over four equivalent Brillouin zones covered by the central sections of the MAPS detector bank (Further, to increase the statistical quality of the data, we averaged the highly similar 35 K and 40 K data. This average is henceforth referred to as 38 K data). Most prominent in the data obtained at 6 K are dispersive spin excitations emerging from $(\pi, \pi)$ and reaching a maximum energy around $14.5$ meV at $(\pi/2, \pi/2)$. The dispersion along the magnetic zone boundary connecting $(\pi, \pi/2)$ and $(\pi/2, 0)$, which was originally reported in Ref. \cite{PhysRevLett.87.037202}, is immediately evident in Fig.\ \ref{fig_NS_raw}(a), as is the depression of the magnon intensity \cite{Christensen15264} at $(\pi, 0)$ compared to $(\pi/2, \pi/2)$. In addition to the spin excitations, Fig.\ \ref{fig_NS_raw}(a) also illustrates the contributions from phonon scattering. Notably, flat bands of phonon scattering with weakly momentum-dependent intensities are observed centered at 7.5 meV and 20 meV. 

Comparing Figs.\ \ref{fig_NS_raw} (a) and (b), it is evident that the dispersion of the spin excitations is similar at 6 K, below $T_N$, and at 20 K, just above $T_N$. This is consistent with earlier reports \cite{birgeneau69BC,PhysRevB.3.1736} for the $S=1$ system K$_2$NiF$_4$. 
A slight broadening is discernible at low energy transfers, whereas the main effect along the magnetic zone boundary is a an overall decrease of intensity. At higher temperatures (Figs.\ \ref{fig_NS_raw}(c) and (d)), the high-energy spin excitations lose definition. The long wavelength excitations appear further broadened compared to the 20 K data, and the phonon scattering is seen to increase, following the Bose occupation factor.  

\begin{figure}[H]
    \centering
    \includegraphics[width=0.8\textwidth]{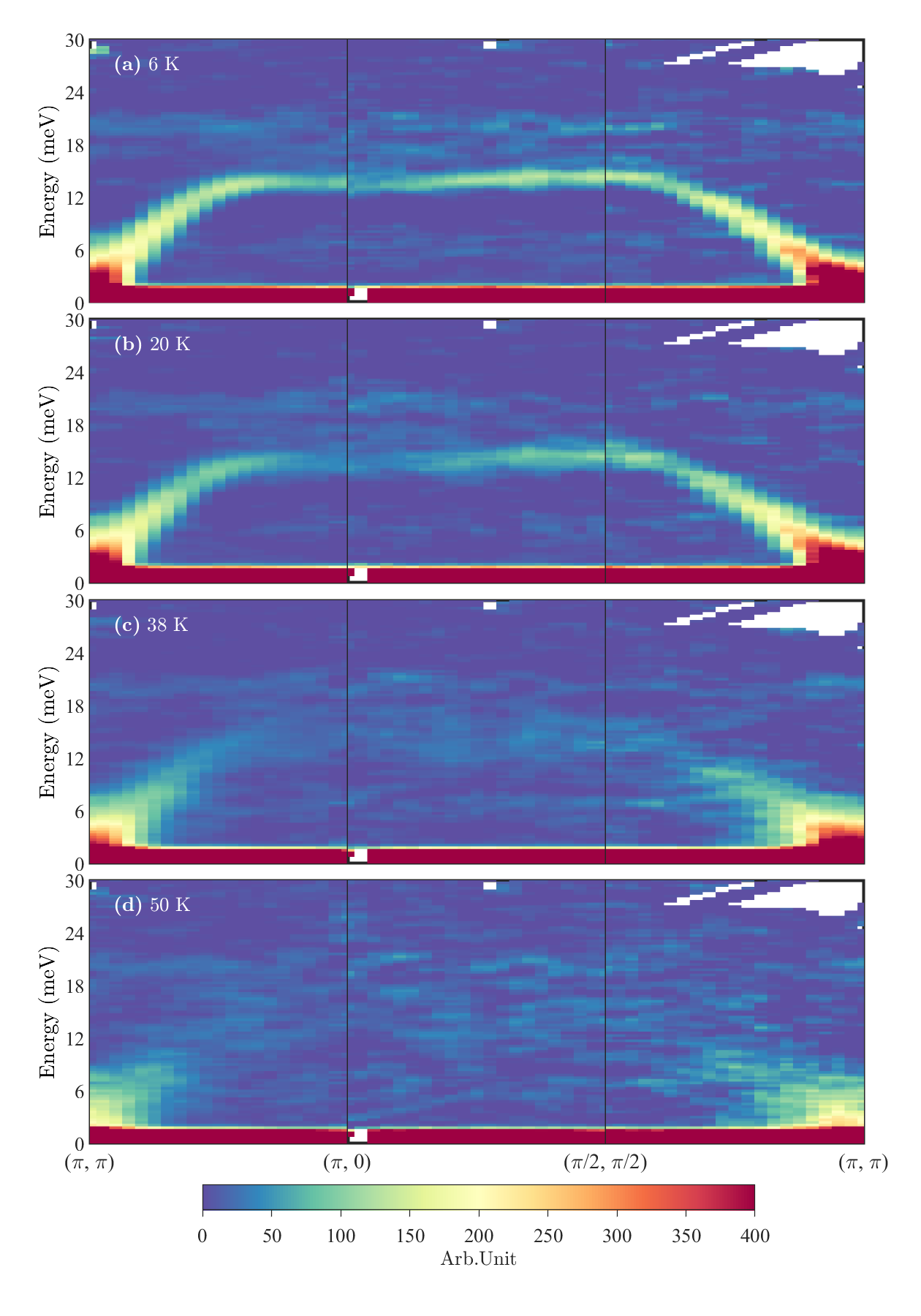}
    \caption{Magnetic excitation spectra, S($\bf{Q}$,$\omega$), of CFTD for temperatures (a), 6~K. (b), 20~K, (c), 38~K and (d), 50~K after subtracting the 80~K spectrum corrected for the Bose occupation factor.}
    \label{fig_BS_80K}
\end{figure}

The evolution of the spin excitations beyond 20 K becomes clearer after subtraction of the 80 K spectrum (which displays negligible magnetic scattering away from $(\pi, \pi)$), from the spectra obtained at lower temperatures. The result of the subtraction procedure is shown in Fig.\ \ref{fig_BS_80K}. While it is clear that the subtraction procedure has not completely eliminated the phonon contributions near $7.5$ meV and $20$ meV, the remnant contributions appear as easily identifiable, weak $Q$-independent contributions at these energies that do not impair the analysis to be presented in the following. (Several other phonon subtraction procedures were tried -- all leading to similar results). Figure\ \ref{fig_BS_80K} (a) and (b) illustrate the same observations made above in connection with Fig.\ \ref{fig_NS_raw}(a) and (b), i.e. neither the existence of a zone boundary dispersion nor that of an intensity decrease at $(\pi, 0)$ compared to $(\pi/2, \pi/2)$ are dependent on the sample being in a magnetically ordered state or at a temperature just above $T_N$. As we continue to heat the sample (Figs.\ \ref{fig_BS_80K}(c) and (d)) the high-energy excitations, which were difficult to identify against the significant phonon background in the raw data shown in Fig.\ \ref{fig_NS_raw}, are now visible as a broad diffuse contribution.

\begin{figure}[H]
    \centering
    \includegraphics[width=0.8\textwidth]{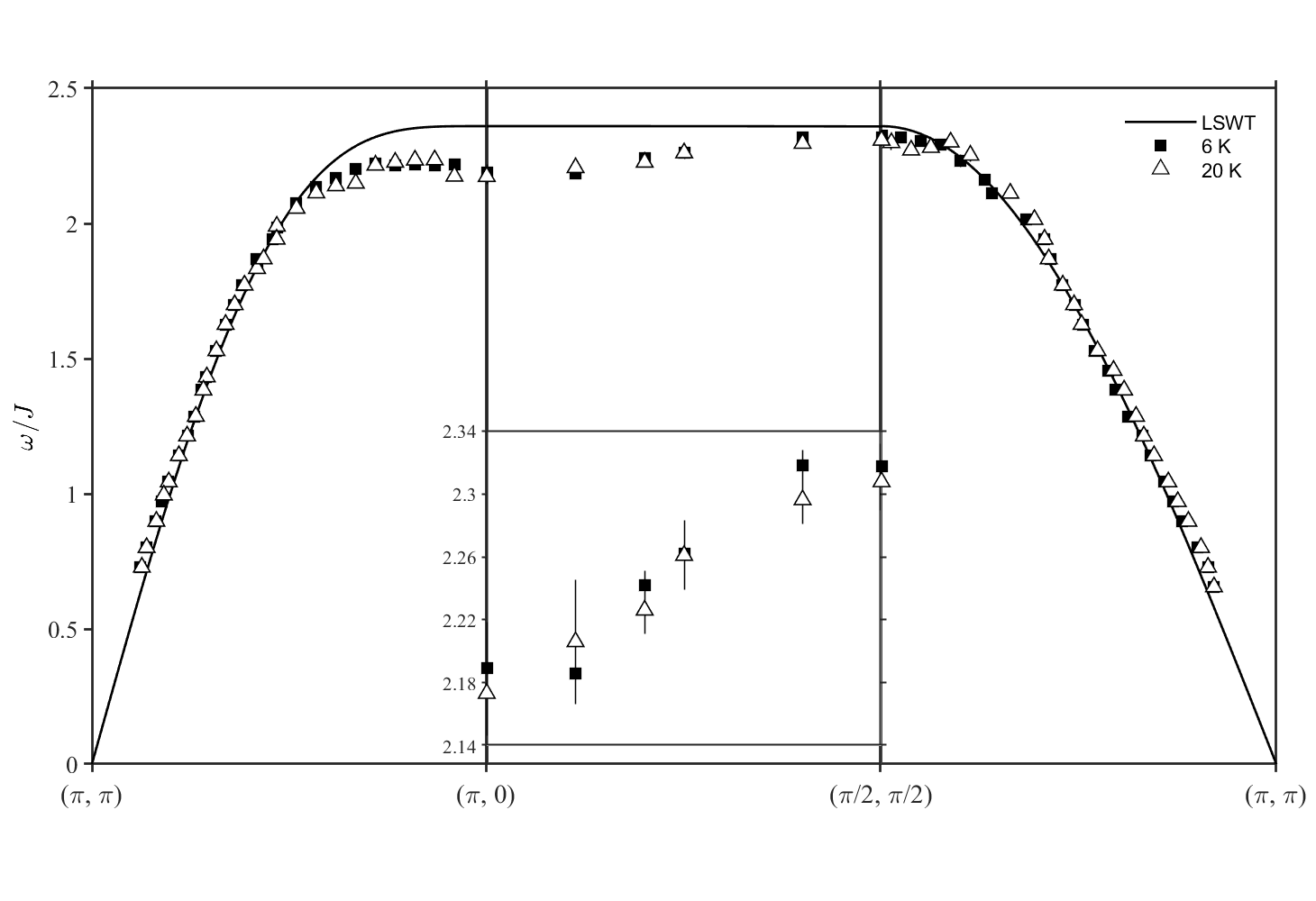}
    \caption{The extracted dispersion of the magnetic excitations along the high-symmetry directions in the Brillouin zone at 6~K (filled squares) and 20~K (open triangles), respectively. The inset shows the detailed dispersion along the zone boundary. The solid line represents linear spin wave theory with $J=6.19$ meV.}
    \label{fig_disp_fit}
\end{figure}

To quantify this evolution of the dispersion of the spin excitations through the ordering transition, the data in Figs.\ \ref{fig_NS_raw} (a)-(b) were analyzed by fitting constant energy cuts (for the low-energy spectrum) and constant momentum cuts (for the high-energy spectrum) to Gaussian lineshapes. Figure\ \ref{fig_disp_fit} shows the results of the fitting procedure. The general observation to make is that the magnon dispersion  -- here plotted as the excitation energy normalized by the exchange constant $J$ --  are found to be essentially indistinguishable at 6 K and 20 K.  

\subsection{Thermal evolution of the zone boundary spectra}
\begin{figure}[H]
    \centering
    \includegraphics[width=0.85\textwidth]{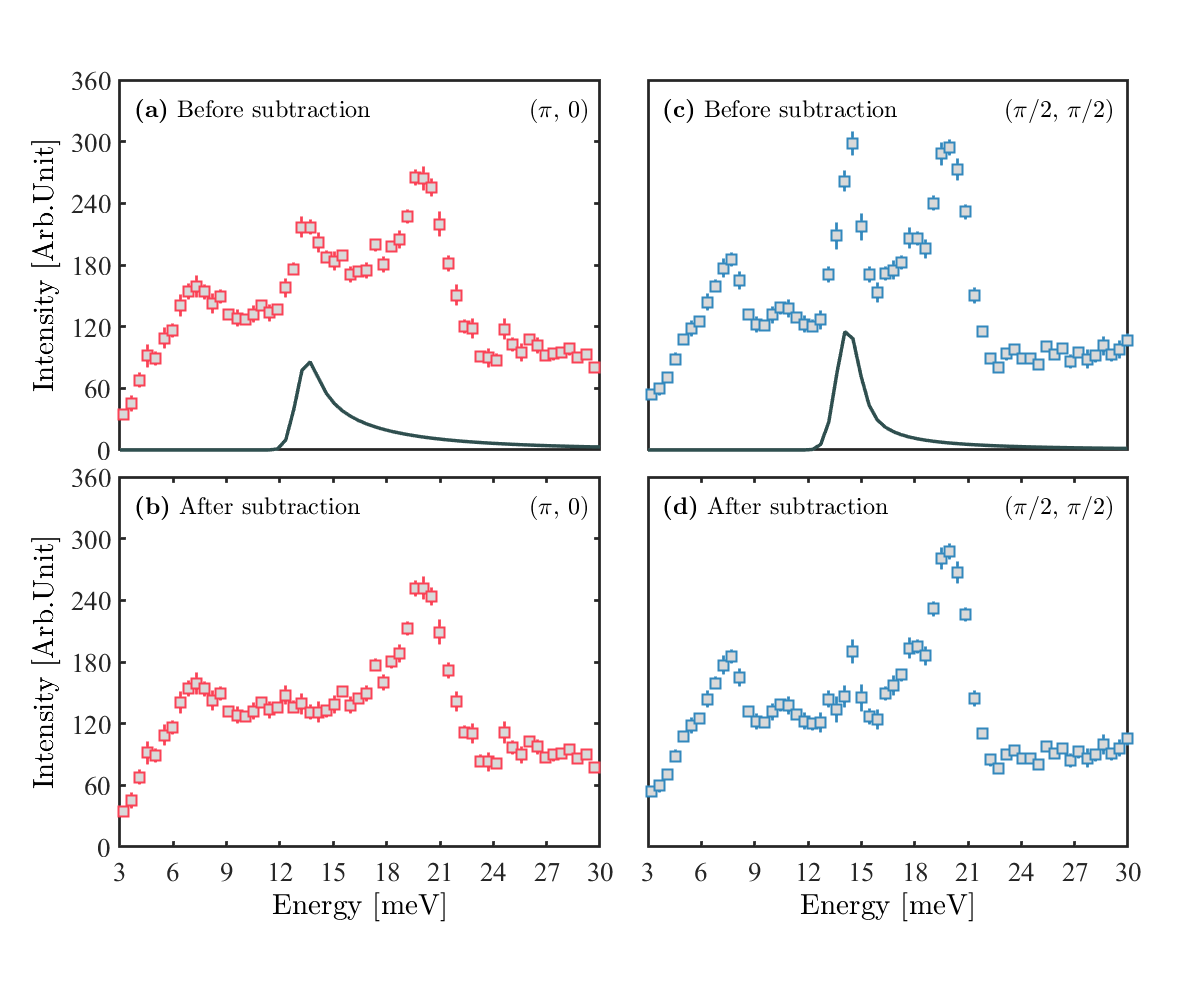}
    \caption{Subtraction of the extrapolated polarized neutron data data from the MAPS 6~K data. Panels (a) and (c) show raw data at $(\pi, 0)$ and $(\pi/2, \pi/2)$, respectively, along with the corresponding magnetic scattering lineshapes extracted from the polarised neutron data in Ref. \cite{dalla2015fractional}. Panels (b) and (d) show the non-magnetic background spectra at the same wavevectors, obtained by subtraction of the lineshape from the data in (a) and (c). After rescaling by the appropriate Bose occupation factors, the spectra in (b) and (d) were subtracted from the raw zone boundary spectra at 20, 38, 50 and 80 K to yield the results in Fig. \ref{Fig5}.}
    \label{Fig4}
\end{figure}

In order to investigate the evolution with temperature of the anomalous zone-boundary spectrum at ($\pi,0$) relative to that at $(\pi/2, \pi/2)$ a different phonon subtraction method beyond a simple
scaled subtraction of the high-temperature (80~K) data had to be developed. The reason is the rapid diminution with increasing temperature of the zone-boundary intensity with temperatures above 20~K (see Fig.\ \ref{fig_BS_80K}).
Several alternative methods were attempted, all yielding qualitatively similar  results. The method found to be most accurate -- as judged by the best removal of the phonon scattering -- was a hybrid one that made use of 
our knowledge from a previous polarized neutron scattering experiment in which we carefully isolated the low-temperature magnetic spectra at the two wavevectors in question from the phonon background. These spectra (transverse plus longitudinal 
response; Figs. 2(a) and (e) in Ref. \cite{dalla2015fractional}) were fitted to phenomenological line shapes, which are indicated by the solid lines in Figs.\ \ref{Fig4}(a) and (c). The line shapes were then subtracted from the 6~K MAPS data set to produce the effective non-magnetic 
background spectra shown in Figs.\ \ref{Fig4}(b) and (d). It is worth noting that apart from the intense phonon bands near 7.5 and 20 meV, the nonmagnetic background at $(\pi/2, \pi/2)$ includes a weaker lattice contribution near 15 meV which was also observed in our polarised neutron experiment, thereby further validating the accuracy of the subtraction procedure. Finally, the non-magnetic background spectra (scaled appropriately by the Bose phonon population factor) were subtracted from the raw 20, 38, 50 and 80~K data sets. 

\begin{figure}[H]
    \centering
    \includegraphics[width=0.85\textwidth]{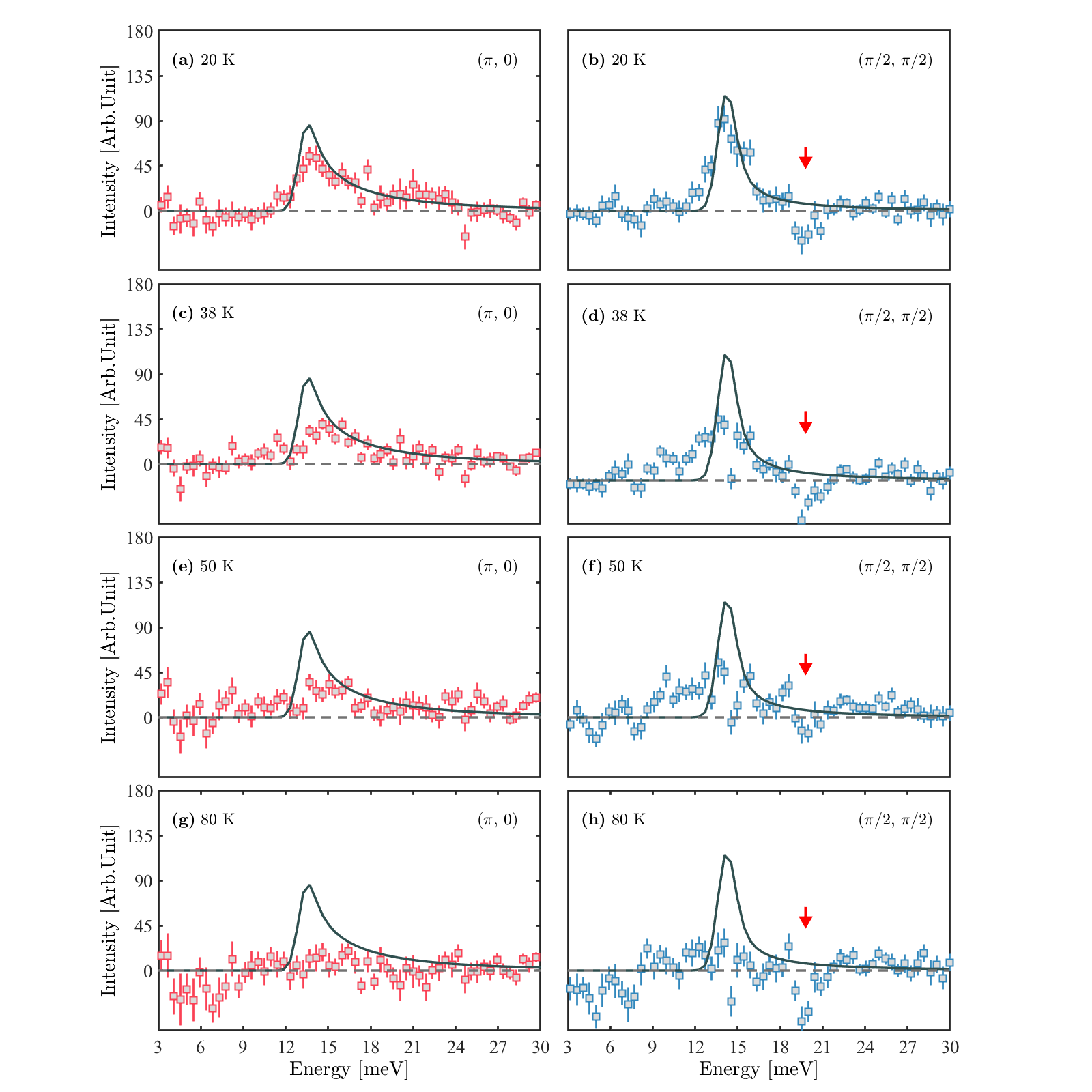}
    \caption{Energy dependence of the measured total magnetic dynamical structure factors at $\bf{Q}$ = $(\pi,0)$ and $\bf{Q}$ = $(\pi/2, \pi/2)$ for different temperatures (a)-(b), 20~K, (c)-(d), 38~K, (e)-(f), 50~K, and (g)-(h), 80~K. The solid lines represent the lineshape of the polarized neutron data from Ref. \cite{dalla2015fractional} which were obtained at 1.5 K. The red arrows indicate positions of an over-subtracted phonon mode.}
    \label{Fig5}
\end{figure}

The spectra resulting from the subtraction method described above are shown in Figs.\ \ref{Fig5}. Inspection of the 80~K data (Figs.\ \ref{Fig5}(g)-(h)) reveals the strengths and weaknesses of the subtraction method. At this temperature, where the magnetic response at high energy is expected to be weak, the method yields an essentially flat response at $(\pi, 0)$ over the energies of interest, in line with expectations. However, it is clear that at $(\pi/2, \pi/2)$ the method overestimates the phonon contribution somewhat for energies below $\sim$9~meV as well as around 20 meV, leading to negative intensities. 
Nonetheless, we consider the method to be robust enough to allow us to make firm deductions on the evolution  of the magnetic spectra at
the two wavevectors for temperatures lower than 80~K.
The data shown in Fig. \ref{Fig5} establish that the zone-boundary magnetic scattering persists up to 50 K $\simeq 3T_N$, but has vanished at 80 K$\simeq 5T_N$. 
We note in particular that 6 K and 20 K zone boundary spectra are similar. This was already evident from the color plots in Figs. \ref{fig_NS_raw}(a)-(b) and \ref{fig_BS_80K}(a)-(b), and revealed by our fits in Fig. \ref{fig_disp_fit}, but is here apparent from a comparison between the data and the 6 K polarised neutron lineshape (black lines in \ref{Fig5}(a)-(b)). At higher temperatures, the zone boundary magnetic scattering continues to weaken. At $(\pi/2,\pi/2)$, the data in Figs. \ref{Fig5}(d) and (f) further reveal a downward energy shift and a substantial broadening towards lower energy transfers.

\begin{figure}[H]
    \centering
    \includegraphics[width=0.8\textwidth]{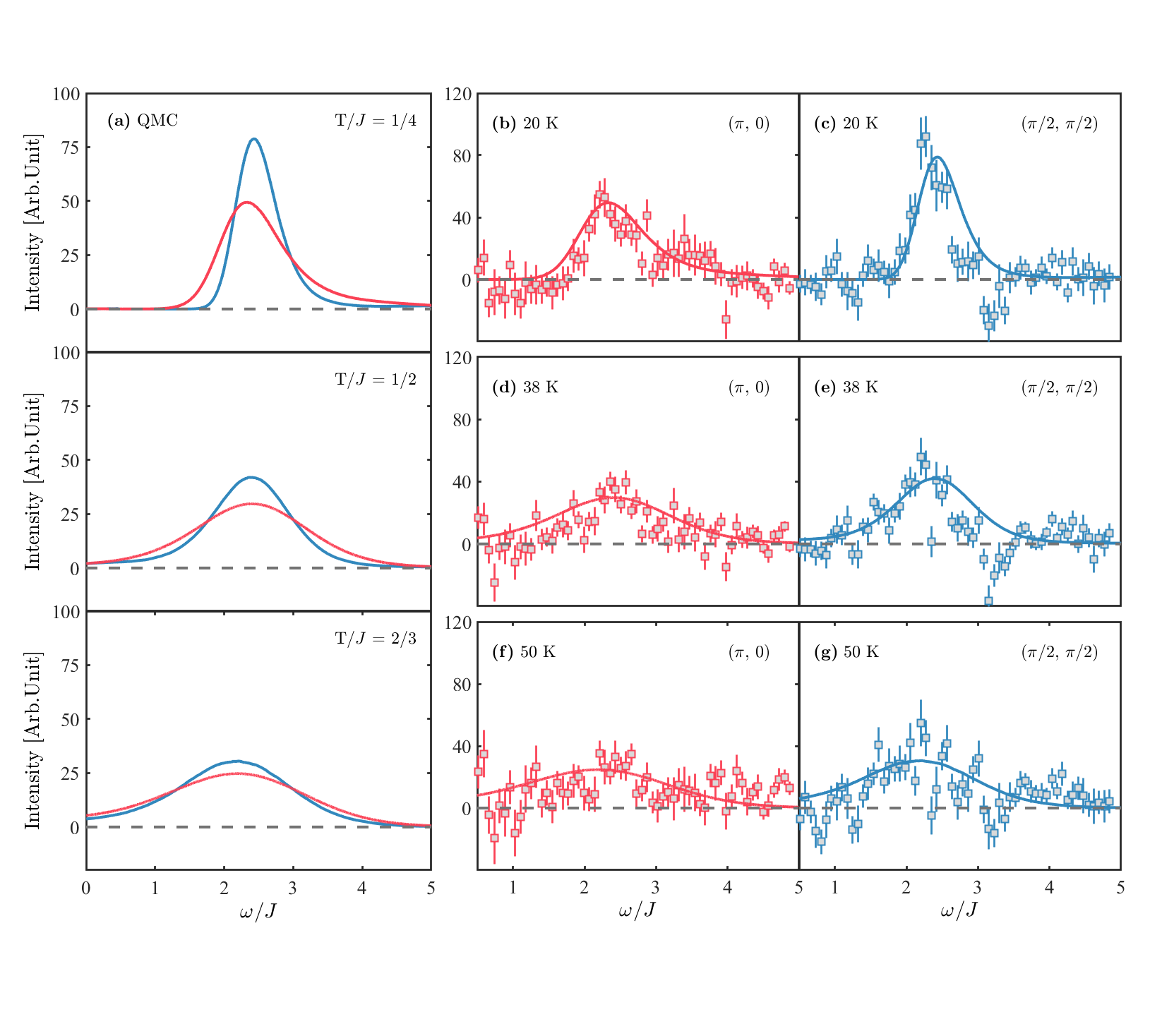}
    \caption{(a) Quantum Monte Carlo calculations of the excitation spectra at $(\pi/2 ,\pi/2)$ (blue lines) and $(\pi ,0)$ (red lines) at $T/J=1/4$, $1/2$ and $2/3$, respectively. With the exchange constant $J=6.19$ meV, relevant for CFTD, this corresponds to temperatures 18, 36 and 48~K. (b)-(g), Comparison of the QMC dynamical structure factors with the experimental data at (b)-(c) 20 K, (d)-(e) 38~K and (f)-(g) 50~K for both $(\pi,0)$ and $(\pi/2, \pi/2)$.}
    \label{Fig6}
\end{figure}

The results of our QMC calculations of $S(\mathbf{Q},\omega)$ are summarised in Fig.\ \ref{Fig6}(a) (Here the $T/J$ values of 1/4, 1/2, 2/3 used for the QMC  calculations correspond to temperatures of 18, 36 and 48~K, respectively, very close to the temperatures used in our experiment). With increasing temperature, $S(\mathbf{Q},\omega)$ computed from QMC is seen to weaken, soften, and broaden to become more symmetric with a loss of the sharper leading edge to lower energies. Of significance is the fact that increasing the temperature to $T$ of order $J$ acts to equalise the response at the two wavevectors i.e. to remove the zone boundary anomaly. 
In Figs.\ \ref{Fig6}(b)-(g) we compare the experimental data to the QMC lineshapes. It is evident that there is good overall agreement, although there is a marked tendency that the QMC calculations predict a broader response at the highest temperature than is observed experimentally. Further theoretical and experimental work will be needed to understand this discrepancy. 

\subsection{Temperature dependence of the zone-centre dispersion}

\begin{figure}[H]
    \centering
    \includegraphics[width=0.75\textwidth]{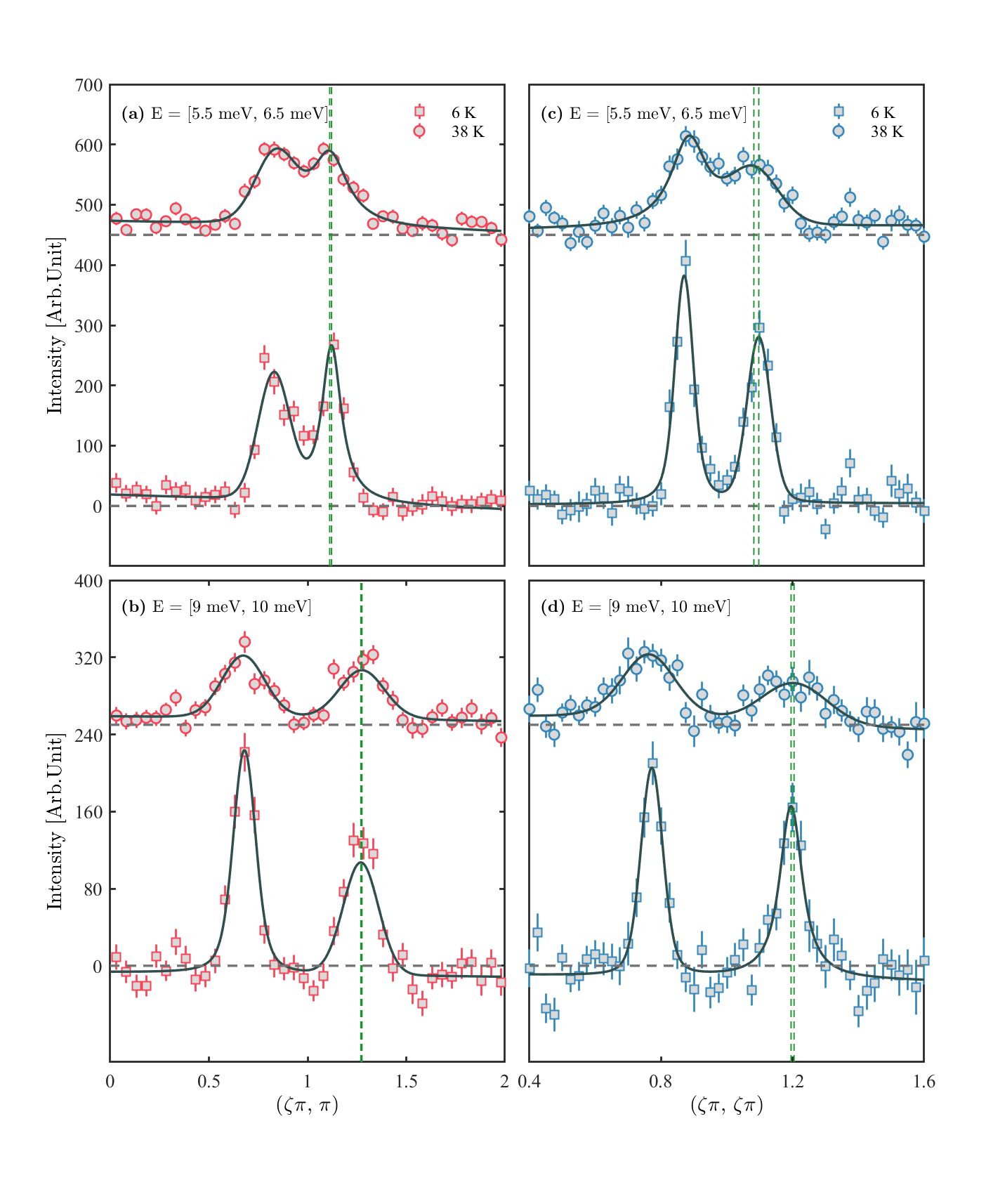}
    \caption{Constant energy cuts through the 80~K background subtracted data at 6~K (squares) and 38~K (circles) seen in Figs. \ref{fig_BS_80K}(a) and (c). The cuts in panels (a) and (c) correspond to the energy range 5.5 to 6.5 meV, while those in (b) and (d) correspond to the range 9 to 10 meV. In the left hand and right hand panels, the cuts through $(\pi, \pi)$ run along the $(1, 0)$ and $(1, 1)$ directions respectively. 
    In all panels the solid lines represent fits to a lineshape consisting of two Voigt functions and the green dashed lines indicate the fitted peak positions at 6 K and 38 K.}
    \label{Fig7}
\end{figure}

Having described the evolution of the zone boundary spectra with temperature, we conclude by analyzing the temperature dependence of the spin excitations dispersing away from $(\pi, \pi)$ along the $(1, 0)$ and $(1, 1)$ directions. Figure \ref{Fig7} compares representative constant energy cuts through the 80 K subtracted data at 6 K and 38 K (Figs. \ref{fig_BS_80K}(a) and (c), respectively). Panels (a) and (b) display cuts along the $(1, 0)$ direction averaged over the energy ranges 5.5-6.5 meV, and 9-10 meV, respectively. In panels (c) and (d) the cuts run along the $(1, 1)$ direction and the energy ranges are the same as in (a) and (b). In all cases, the solid lines represent fits to two Voigt lineshapes for the two counter-propagating spin excitation modes. The fits illustrate a clear broadening of the excitations with temperature, as expected, but little or no indication of changes in their wavevector.

\begin{figure}[H]
    \centering
    \includegraphics[width=0.75\textwidth]{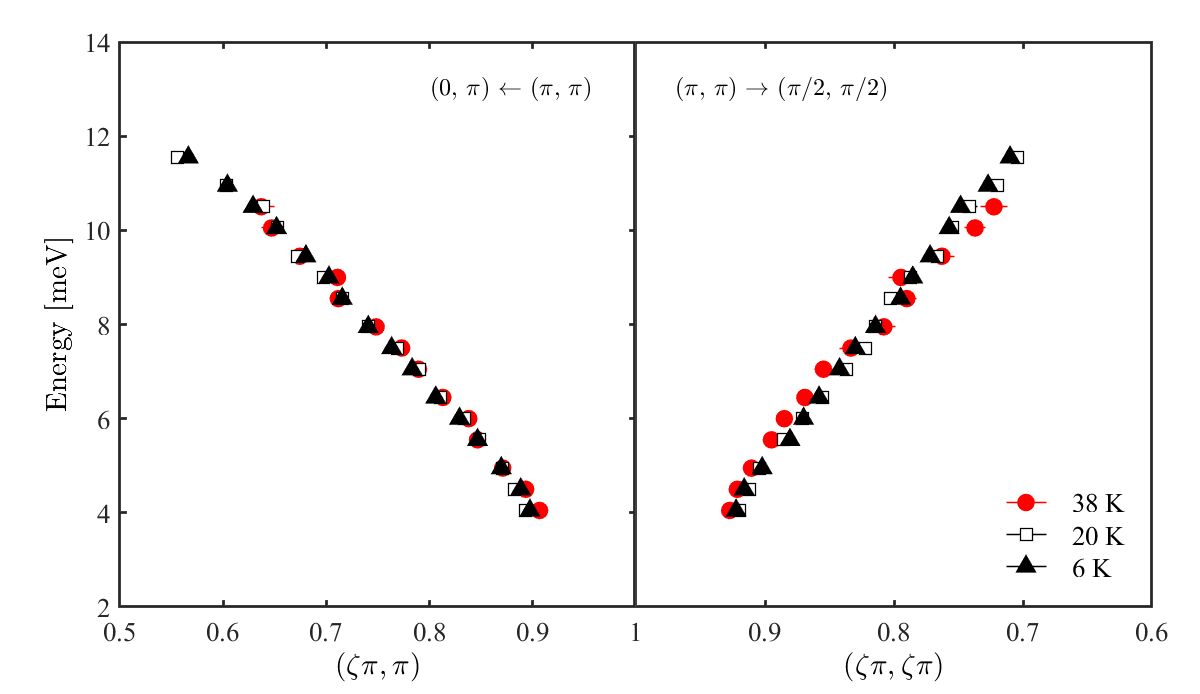}
    \caption{Dispersion of the low energy spin excitations at 6~K (black triangles), 20~K (open squares), and 38~K (red cicles), respectively, obtained from fits to cuts such as those in Fig. \ref{Fig7}.}
    \label{Fig8}
\end{figure}

Figure \ref{Fig8} summarises the dispersion of the low-energy excitations between 4 meV and 11.5 meV (We are not able to reliably obtain information about the lowest energy excitations after subtraction of the 80 K data set, since the latter contains a broad remnant magnetic contribution below 4 meV) at 6 K, 20 K and 38 K. Overall, the results show little change with temperature, except for weak indications of hardening (softening) at low (high) energies for spin excitations dispersing along the $(1, 1)$ direction.

\section{Discussion and Conclusions}

Our analysis of the thermal evolution of the spin excitations in the two-dimensional quantum Heisenberg antiferromagnet CFTD has revealed several novel results: (i) The spin excitation spectra at 6 K and 20 K are remarkably similar. In particular, the $(\pi, 0)$ anomaly observed in the magnetically ordered state remains at 20 K. (ii) Further heating causes the zone boundary magnetic scattering to gradually broaden and lose spectral weight, but it remains visible at temperatures $T/J \simeq 2/3$. (iii) As it broadens with temperature, the lineshape at $(\pi,0)$ becomes more symmetric but only near $T\simeq J$ does the difference between $(\pi,0)$ and $(\pi/2,\pi/2)$ fully disappear. These observations are broadly consistent with our QMC calculations. (iv) Tracking the lower-energy spin excitations shows little evidence for changes in the dispersion up to to $T/J \simeq 1/2$. In the following, we proceed to discuss these results.

Our study of zone boundary excitations in the 2DQHAFSL shines light on the interplay between quantum and thermal effects in quasi-two-dimensional Heisenberg antiferromagnets. Naively, the persistence of relatively well-defined zone-boundary excitations at $\omega\!\approx\!2J$ up to $T\!\approx\!2/3J$, a temperature for which spin correlations are very short-ranged \cite{PhysRevLett.82.3152}, lends itself to a simple interpretation in real-space. In an instantaneous picture, a given spin remains surrounded by four anti-aligned nearest-neighbors, such that the cost of a spin flip remains $\approx 2J$. In other words, the energy of zone-boundary excitations does not scale with the nearest-neighbor expectation value $\langle {\bf S}_i \cdot {\bf S}_j \rangle$ but instead simply reflects the energy of localized spin-flips. The similarity of the $(\pi/2,\pi/2)$ and $(\pi,0)$ lineshapes above $T = 2J/3$ provides further support for this naive picture because it indicates that lattice-directionality effects no longer matter at elevated temperature. However, beyond this simple picture, we note that our broadened zone-boundary excitations are centered at an energy $\omega > 2J$ even as $T = 2J/3$. In the zero-temperature limit, the enhanced bandwidth of spin wave excitations (compared to $2J$) directly stems from quantum corrections to the excitation energies. Our results therefore open the intriguing possibility that quantum renormalization effects remain active at elevated temperatures.

We believe that the robustness of the $(\pi, 0)$ anomaly in CFTD above the N\'eel ordering temperature and its gradual evolution with temperature provides an important benchmark for future work aiming at testing the validity of a theoretical description based on spin-waves, spinons and/or proximity to a deconfined phase in the 2DQHAFSL. In that regard, it may be useful to carry out $T>0$ QMC calculations of $S({\bf Q},\omega)$ also for the $J$-$Q$ model studied at $T=0$ in Ref. \cite{PhysRevX.7.041072}. In this model multi-spin interactions compete with the Heisenberg  exchange and eventually lead to a deconfined quantum critical point. At $T=0$, the $(\pi, 0)$ anomaly strengthens rapidly with increasing multi-spin interaction. Monitoring $S({\bf Q},\omega)$ for both ${\bf Q}=(\pi, 0)$ and $(\pi/2, \pi/2)$ as the critical point is approached should shed additional light on the specific spectral signatures of spinon deconfinement at elevated temperatures.

More broadly, the issue of the applicability and limitations of spin-wave theory to describe the magnetic excitations of quasi-two-dimensional insulating spin systems has recently witnessed a renewed interest. Current discussions include the existence of fractionalized excitations in the Kitaev-Honeycomb magnet $\alpha$-RuCl3 \cite{Banerjee2017,Winter2017,Samarakoon2018} and the nature of magnetic excitations at the M-point of the Brillouin zone in the triangular-lattice Heisenberg antiferromagnet \cite{Mourigal2013,Ghioldi2018,Verresen2019,Ferrari2019} as exemplified by the compound Ba$_3$CoSb$_2$O$_9$ \cite{Ma2016,Ito2017}. By extending our study of the zone boundary excitations of the 2DQHAFSL to finite temperatures, we hope to pave the way for future neutron scattering and numerical studies aiming at understanding the interplay between quantum and thermal effect in spin systems. 

It is worthwhile considering differences and similarities between the classical spin HAF Rb$_2$MnF$_4$ ($S=5/2$) studied by Roger Cowley and his collaborators \cite{huberman2008} and the quantum HAF CFTD investigated here. In the former, it was found that for temperatures greater than $T_N$ but smaller than the Curie-Weiss temperature, excitations with wavelengths shorter than the inverse correlation length remain well-defined in the sense that their energy width $\Gamma$ is insignificantly compared to $\hbar\omega$. The present study of an ideal 2DQHAFSL is not as detailed as Ref. \cite{huberman2008}, and does not extend to the Curie-Weiss temperature, but nevertheless indicates that the intermediate and high-energy spin excitations of CFTD are less prone to softening than in the classical analogue. In particular, in Rb$_2$MnF$_4$, the zone boundary excitation energy at $(\pi, 0)$ is broadened and renormalized downwards by about 7-10$\%$ already at $T_N$, while in CFTD, we observe no renormalization at 20 K, while the peak intensity of the sharp part of the magnetic spectrum is suppressed. It is plausible that the cause of these differences are ultimately a result of the different ratios $f=T_N/\hbar\omega_{ZB}(T=0)$ in Rb$_2$MnF$_4$ ($f \simeq 0.5$) and CFTD ($f \simeq 0.1$), as considered already in Refs. \cite{PhysRevB.3.1736,Silberglitt71}. 

Finally, in the broader context of quantum materials and strongly correlated electron systems, it is interesting to compare the present observations of how the excitation spectrum evolves as thermal fluctuations destroy antiferromagnetic correlations to the situation in doped cuprates, where doping with mobile holes destroy the antiferromagnetic order. With the advent of RIXS doping dependent studies have revealed that broadened, dispersive excitations along the antinodal $(1, 0)$ direction survive to surprisingly high doping values \cite{LeTacon2011,Dean2013}. This is consistent with our current observation in CFTD. However in RIXS measurements on Bi$_2$Sr$_2$CaCu$_2$O$_{8+\delta}$, it was found that the dispersion along the nodal $(1, 1)$ direction softened dramatically even for underdoped samples \cite{guarise2014}. This is in strong contrast to our observation of temperature effects in an undoped system, where the excitations remain with little softening along both directions. 

\ack
We are delighted by this opportunity to contribute to this special issue of JPMC celebrating the life and works of Roger Cowley. Roger's sublime talent enabled him to produce seminal contributions to a staggering array of subjects, many of which are documented in this edition. However, in our opinion these accomplishments form only part of his enduring legacy. Of equal importance is the manner in which he engaged with other researchers, and most especially the great kindness he showed to those embarking on their careers. For those working in his research group he created an environment that was both supportive and encouraging of independent initiative and adventure.  To the many others he encountered at the various National Laboratories and central facilities where he worked throughout his career, he offered his unflagging interest and unique insight. And last but not least, there was an implied imperative to enjoy the experience: for Roger, there was no situation  -- failing experiment, intractable calculation, etc. -- that could not be improved by a good competitive game of tennis, hike or the unarmed combat to be found on a  croquet lawn.

Work in London was supported by the Engineering and Physical Sciences Research Council (Grant number EP/P013449/1). Work in Denmark was supported by the instrument center Danscatt funded by the Danish Agency for Science, Technology, and Innovation. AWS was supported by the NSF under Grant No. DMR-1710170 and by a Simons Investigator Grant. The QMC calculations were carried out on Boston University's Shared Computing Cluster. 
 
\section*{References}

\bibliographystyle{iopart-num}
\bibliography{CFTD_finiteT}

\end{document}